\documentclass[letterpaper, 10 pt, conference]{ieeeconf}  

\IEEEoverridecommandlockouts                              

\usepackage{mathtools,amsmath,amssymb}
\usepackage{color,soul}
\usepackage{comment}
\newtheorem{remark}{Remark}
\newtheorem{theorem}{Theorem}

\begin{document}

\title{\LARGE \bf Data--driven predictive control with estimated prediction matrices and integral action}

\author{P. C. N. Verheijen, G. R. Gon\c{c}alves da Silva and M. Lazar
\thanks{The authors are with the Department of Electrical Engineering, Eindhoven University of Technology, The Netherlands: 
{\tt\small p.c.n.verheijen@student.tue.nl, g.goncalves.da.silva@tue.nl, m.lazar@tue.nl}}%
}

\maketitle
\pagestyle{empty}
\begin{abstract}%
This paper presents a data--driven approach to the design of predictive controllers. The prediction matrices utilized in standard model predictive control (MPC) algorithms are typically constructed using knowledge of a system model such as, state--space or input--output models. Instead, we directly estimate the prediction matrices relating future outputs with current and future inputs from measured data, off--line. On--line, the developed data--driven predictive controller reduces to solving a quadratic program with a similar structure and complexity as linear MPC. Additionally, we develop a new procedure for estimating prediction matrices from data for predictive controllers with integral action, corresponding to the rate--based formulation of linear MPC. The effectiveness of the developed data--driven predictive controller is illustrated on position control of a linear motor model.   %
\end{abstract}


\section{Introduction}
Data--driven (DD) control design methods have recently gained an increased attention from the control systems community due to the digital technology trends involving big--data and artificial intelligence systems, see, for example, \cite{Controlforfuture:2017,Markovsky&Rapisarda:2008} and the references therein. The main idea of data--driven control is to eliminate the standard separation in model--based controller design, i.e., first obtain a system model, by identification or first principles, and then design a controller. Instead, in DD controller design, more freedom is allowed in mixing identification (or estimation) with controller design, and even direct controller synthesis from data. 

Much of the existing data--driven methods for controller design, see, e.g., \cite{Bazanella&Campestrini&Eckhard:2012,Hou&Wang:2013,Campi&Lecchini&Savaresi:2002,Campestrini:Eckhard:Bazanella:Gevers:2017} and the references therein, have originated within the field of adaptive control and make use of the so--called model reference control framework. 
In these approaches, instead of first deriving a system model, measured input--output (I/O) data is directly used to identify a parameterized \textit{controller} chosen \textit{a priori}, given a performance requirement, i.e., the reference model. Since it is hard to guarantee closed--loop stability while tuning the controller with these methods without largely compromising performance, an \textit{a posteriori} data--driven certification procedure has also been reported in \cite{Goncalves:2020}.

Recently, these ideas have been extended to other types of robust control problems, such as state--feedback stabilization \cite{Waarde&Eising&Trentelman&Camlibel:2020}, linear quadratic regulator (LQR) design \cite{LQRpredict,Goncalves&Bazanella&Lorenzini&Campestrini:2018}, and output feedback with an embedded prediction step \cite{Persis&Tesi:2019}. Combining ideas of state--feedback with prediction theory fits well in the framework of model predictive control (MPC) \cite{Rawlings&Mayne&Diehl:2017}, which can also be described by a given performance (whether a reference model or weighting matrices) with a parameterized structure (the prediction matrices, typically constructed using a system model).

MPC approaches based on input--output step--response system models \cite{Clarke&Mohtadi&Tuffs:1987} or finite impulse response (FIR) system models \cite{Prasath&Jorgensen:2008}, have been at the core of the MPC research for a long time. These methods require an identification step to obtain the FIR model (usually by applying some impulses to the system or by differentiating its step response) and an estimation of the system order, and then plugging in these parameters into the prediction matrices. Such an approach, which still resorts to a system identification procedure, is referred to as an \emph{indirect} data--driven approach to controller design.  Another possibility is to combine model--based and data--driven approaches within predictive control algorithms, see, for example, \cite{Carron:2019} and the references therein, where a linear state--space model is used for describing known system dynamics and a learning data--driven approach is used to model unknown disturbances.

\emph{Direct} approaches to data--driven predictive control, inspired by \cite{Markovsky:2006}, were proposed recently in \cite{Yang:2015} and \cite{DataEnabled:2019}. These approaches parameterize future predicted outputs as a linear combination of past inputs and outputs, and future inputs, via Hankel matrices of I/O data and a set of coefficients (parameters). Such methods do not require any off--line identification, at the cost of computing the set of coefficients on--line, simultaneously with computing future control inputs and predicted outputs. This makes the number of optimization variables dependent of both the prediction horizon and the size of the Hankel data matrices. It is worth to mention that \cite{Yang:2015} computes on--line an additive control input that adjusts an unconstrained predictive control law, while \cite{DataEnabled:2019} computes on--line the constrained predictive control input directly, as in standard MPC. Also, \cite{Yang:2015} proposes a method to include integral action, while \cite{DataEnabled:2019} develops solutions for dealing with noisy data. Stability and robustness of Hankel matrices based predictive controllers have been recently studied in \cite{Berberich:2020}. 

Overall, \emph{direct} data--driven predictive controllers are sensitive to noisy data, since they do not inherit any of the standard properties of an intermediate step of estimation (and also due to multiplicative noise terms in the prediction step). Another relevant remark to MPC design \cite{Rawlings&Mayne&Diehl:2017} versus data--driven predictive control (DPC) design is that stability and optimal performance is not anymore easily determined by the cost function tuning. Indeed, while in linear MPC the LQR control law can be recovered if constraints are not active via a suitable terminal penalty, this is no longer the case in DPC, even under the assumption that the full state is measurable.

An unconstrained solution to an input-output data-driven Subspace Predictive Control algorithm has been derived in \cite{favoreel:1999}. However, the authors suggest to compute the prediction matrices by using a pre-step of QR decomposition of the data matrices, and it is not clear how this computation is affected by noise. They then propose a second step to reduce the state prediction matrix to an estimate of the system order. \cite{kadali:2003} builds on the input-output GPC approach, including only upper and lower bound type of constraints and proposing an integral action assuming a particular ARIMAX model structure, so they can take the discrete output difference as feedback. Therein, integral prediction matrices are not directly identified from data, instead, they identify the non-integral prediction matrices first, and then they use summation of the specific terms.
The method developed in this work do not require any type of system order estimate and we formulate it such that inclusion of constraints follow the model-based approach. We also derive the identification procedure for identifying prediction matrices for rate-based integral action state-space MPC with formal guarantees (Theorem 1) without assuming a specific noise model.





Motivated by the current status in DPC design, in this paper we develop an approach to data--driven predictive control that offers an attractive compromise between \emph{indirect} and \emph{direct} approaches to data--driven control. The approach is inspired by the recent data--driven LQR design proposed in \cite{Goncalves&Bazanella&Lorenzini&Campestrini:2018}. Firstly, instead of identifying a state--space or FIR system model and then building the MPC prediction matrices, the developed approach directly estimates the full prediction matrices from measured data. Estimation of the prediction matrices makes the developed approach less sensitive to noisy data. We also show that these matrices are comprised of the system Markov parameters, so they relate to the MPC approaches using FIR models, but without requiring a priori knowledge of the system order. Moreover, we show that if the state sequence can be measured or the prediction horizon is long enough, then one can recover the corresponding LQR controller by using the derived DPC algorithm. Motivated by the need of off-set free control in practice, another contribution of this paper is the derivation of a rate--based DPC algorithm with integral action, similar to linear MPC with integral action \cite{Wang:2009}. The performance and robustness of the developed data--driven predictive controller is illustrated for position control of a linear motor model.







\section{Preliminaries}
\label{sec:2}
In this section we recall the standard linear MPC formulation \cite{Rawlings&Mayne&Diehl:2017} using prediction matrices based on linear discrete--time state--space models, i.e.,
    \begin{equation}
    \begin{aligned}
    \label{eq:sys1}
    x(k+1) &= Ax(k) + Bu(k) \\
    y(k) &= Cx(k)
    \end{aligned}, \quad k\in\mathbb{N},
    \end{equation}
    where $x(k)\in\mathbb{R}^n$ is the state, $u(k)\in\mathbb{R}^m$ is the input, $y(k)\in\mathbb{R}^q$ is the output and $k$ denotes the discrete--time index. Assuming for simplicity 
    that the regulation objective is to control the output to the origin, in MPC one computes a sequence of optimal control inputs $U_k^\ast:=U^\ast(x(k))$ every time instant $k$ by minimizing a cost function, e.g.,
    \begin{equation}
    \label{eq:CostFcn}
    \begin{aligned}
    J(x(k), U_k) &:=y_{N|k}^ TPy_{N|k} + \sum_{i=0}^{N-1}(y_{i|k}^{T}Qy_{i|k} + u_{i|k}^{T}Ru_{i|k}) \\
    &= U_k^T(\Psi+\Gamma^T\Omega\Gamma)U_k + 2U_k^T\Gamma^T\Omega\Phi x(k) + \\
		&~~~~x(k)^T(C^TQC+\Phi^T\Omega\Phi)x(k),
    \end{aligned}
    \end{equation}
   where $U_k:=\{u_{0|k},\ldots,u_{N-1|k}\}$, $x_{0|k}=x(k)$, $u(k)=u_{0|k}$, $x_{i+1|k}=Ax_{i|k}+Bu_{i|k}$ and $y_{i|k}=Cx_{i|k}$. Above $Q$ and $R$ are positive definite symmetric weighting matrices of appropriate dimensions, $N$ is the prediction horizon (we assume that a control horizon equal to the prediction horizon is used for brevity) and $P$ is a terminal weight matrix, which is typically taken equal to the solution of a corresponding discrete--time algebraic Riccati equation (DARE). Moreover, the matrices utilized in \eqref{eq:CostFcn} are defined as follows:
    \begin{equation}
    \begin{aligned}
    \label{eq:MatDef}
    \Phi & = \begin{bmatrix}CA \\ CA^2 \\ \vdots \\ CA^N\end{bmatrix}, ~~ \Gamma = \begin{bmatrix} CB & 0 & \dddot{} & 0 \\ CAB & CB & \dddot{} & 0 \\ \vdots & \vdots & \ddots & \vdots \\  CA^{N-1}B & CA^{N-2}B & \dddot{} & CB \end{bmatrix}, \\
  \Omega & = \text{diag}(Q, \dddot{}, Q, P), ~~~~ \Psi = \text{diag}(R, R, \dddot{}, R).
    \end{aligned}
    \end{equation}
    
    The unconstrained MPC control law (i.e., if output and input constraints are neglected) can be computed by taking the gradient of $J$ with respect to $U_k$ and set it equal to $0$, i.e.:
\begin{equation}
    \begin{aligned}
    \label{eq:OptU}
    \nabla_{U_k}J(x(k), U_k) &= 2(\Psi+\Gamma^T\Omega\Gamma)U_k + 2\Gamma^T\Omega\Phi x(k) = 0, \\
    U_k^*(x(k)) &= -G^{-1}F\Phi x(k), \quad\text{where} \\
    G &= 2(\Psi+\Gamma^T\Omega\Gamma), \quad F = 2\Gamma^T\Omega. \\
    \end{aligned}
    \end{equation}
    When output and input constraints are added to the minimization of $J$, the constrained MPC control law is computed on--line, by solving a quadratic program, i.e.,
	\begin{equation}
	\begin{aligned}
	\label{eq:QP}
	\min_{U_k}& \frac{1}{2}U_k^TGU_k + U_k^TF\Phi x(k) \\
	\text{subject to: }& \mathcal{L}U_k \leq c-\mathcal{D}y(k) - \mathcal{M}\Phi x(k) 
	\end{aligned}
	\end{equation}
	where the derivation of the matrices $\mathcal{L}$, $\mathcal{D}$, $\mathcal{M}$ and the vector $c$ is illustrated next. It is worth mentioning that the terms in \eqref{eq:CostFcn} that do not depend on $U_k$ are omitted in \eqref{eq:QP}, as they do not influence the corresponding optimum.
	
	Consider linear constraints in the outputs and inputs, i.e.
	\begin{equation}
	\begin{aligned}
	\label{eq:constSingle}
	& M_iy_{i|k}+E_iu_{i|k}\leq b_i, ~~ \forall i = 0,1,\dddot{}, N-1, \quad \text{and} \\
	& M_N y_{N|k}\leq b_N,
	\end{aligned}
	\end{equation}
	for suitable matrices $M_i,E_i$ and vectors $b_i$. These constraints can be aggregated as follows
	\begin{equation}
	\begin{aligned}
	\label{eq:constr1}
	\begin{bmatrix}M_0 \\ 0 \\ \vdots \\ 0 \end{bmatrix}y_{0|k}+\begin{bmatrix} 0 & \dddot{} & 0 \\ M_1 & \dddot{} & 0 \\ \vdots & \ddots & \vdots \\ 0 & \dddot{} & M_N \end{bmatrix}\begin{bmatrix}y_{1|k} \\ \vdots \\ y_{N|k}\end{bmatrix} & + \\
		\begin{bmatrix}E_0 & \dddot{} & 0\\\vdots & \ddots & \vdots \\ 0 & \dddot{} & E_{N-1} \\ 0 & \dddot{} & 0 \end{bmatrix}\begin{bmatrix}u_{0|k} \\ \vdots \\ u_{N-1|k}\end{bmatrix}&\leq\begin{bmatrix}b_0 \\ \vdots \\ b_N\end{bmatrix},\quad\text{or} \\
	\mathcal{D}y(k)+\mathcal{M}Y_k+\mathcal{E}U_k &\leq c.
	\end{aligned}
	\end{equation}
	Substituting the future predicted outputs $Y_k = \Phi x(k) + \Gamma U_k$ into equation \eqref{eq:constr1} results in 
	\begin{equation}
	\begin{aligned}
	\label{eq:constr2}
	\mathcal{L}U_k &\leq c - \mathcal{M}\Phi x(k) - \mathcal{D}y(k), ~~ \text{with} ~~ \mathcal{L} = (\mathcal{M}\Gamma+\mathcal{E}).
	\end{aligned}
	\end{equation}
The MPC control law is extracted from $U^\ast(x(k))$ as follows:
	\begin{equation}
	\begin{aligned}
	\label{eq:Optu}
	u^*(x(k)) &= [I_m \quad \mathbf{0}_{(N-m) \times m}]U_k^*(x(k)).
	\end{aligned}
	\end{equation}
Next, we will present a data--driven approach to predictive control design, which is based on estimating the prediction matrix $\Gamma$ and the predicted ``free response'' $\Phi x(k)$ directly from measured input--output (or state) data.


\section{Data--driven predictive control with estimated prediction matrices} \label{sec:3}

Instead of using an identified state--space model to build the prediction matrices $\Phi$ and $\Gamma$ (see \eqref{eq:OptU} and \eqref{eq:constr2}), we aim to directly compute these matrices from I/O data. The data can either be collected whilst controlling the system on--line or obtained off--line from an experiment on the system. Even though collecting data on--line is preferable, as no prior experiment is required, there is no guarantee of persistence of excitation for the inputs generated by a controller. This can yield not only ill--conditioned prediction matrices, but it can strongly degrade initial performance due to the short measurement horizon. Therefore, in this work we assume the data is measured off--line.

\subsection{Estimating the prediction matrices}\label{ssec:estimation}

We start by showing how to obtain $\Gamma$ from measured data, using an ARMarkov model as reported in \cite{Goncalves&Bazanella&Lorenzini&Campestrini:2018,LQRpredict,armakov}. Suppose the input and output vectors are structured as (for any $k\geq 0, N \geq 1$):

	$$\mathbf{y}(k) = \begin{bmatrix}y(k) \\ y(k+1) \\ \vdots \\ y(k+N-1) \end{bmatrix}, \quad \mathbf{u}(k) = \begin{bmatrix}u(k) \\ u(k+1) \\ \vdots \\ u(k+N-1) \end{bmatrix}.$$
	These vectors can be stacked in the following Hankel matrices:
	\begin{equation}\label{eq:hankel_data}
    \begin{aligned}
	\mathbf{Y}_p &= \begin{bmatrix}\mathbf{y}(1) & \dddot{} & \mathbf{y}(L+1) \end{bmatrix}, \\
	\mathbf{U}_p &= \begin{bmatrix}\mathbf{u}(0) & \dddot{} & \mathbf{u}(L) \end{bmatrix}, \\
	\mathbf{Y}_f &= \begin{bmatrix}\mathbf{y}(N+1) & \dddot{} & \mathbf{y}(N+L+1) \end{bmatrix},\\
	\mathbf{U}_f &= \begin{bmatrix}\mathbf{u}(N) & \dddot{} & \mathbf{u}(N+L) \end{bmatrix}, 
	\end{aligned}
    \end{equation}
	representing the so--called ``past'' and ``future'' I/O data. Above $L$ represents some chosen measurement horizon, which we will define later. 
	According to \cite{Goncalves&Bazanella&Lorenzini&Campestrini:2018,SubspaceIDOverschee&DeMoor}, the relation between these matrices can be described as follows:
	\begin{equation}
	\label{eq:Hrel}
	\begin{aligned}
	\mathbf{Y}_p &= \Phi X_p+\Gamma\mathbf{U}_p, \\
	\mathbf{Y}_f &= \Phi X_f + \Gamma\mathbf{U}_f, \\
	X_f & = A^{N}X_p+D\mathbf{U}_p,
	\end{aligned}
	\end{equation}
where $X_p = \begin{bmatrix} x(0) & x(1) & \dddot{} & x(L) \end{bmatrix}, ~X_f = \begin{bmatrix} x(N) & x(N+1) & \dddot{} & x(N+L) \end{bmatrix}$, and $D$ is the extended controllability matrix
	$D = \begin{bmatrix} A^{N-1}B & \dddot{} & AB & B \end{bmatrix}$.
	According to \cite{armakov}, as long as $(N+1)q\geq n$, it is guaranteed for an observable system that there exists a matrix $M$ such that $A^{N} + M\Phi = 0.$ 
	This allows us to rewrite $\mathbf{Y}_f$ without state information:
	\begin{equation*}
	\begin{aligned}
	\label{eq:DDid1}
	\mathbf{Y}_f &= \Phi A^{N}X_p+\Phi D\mathbf{U}_p+\Gamma\mathbf{U}_f \\
	&= -\Phi M\Phi X_p+\Phi D\mathbf{U}_p+\Gamma\mathbf{U}_f \\
	&= -\Phi M(\mathbf{Y}_p -\Gamma\mathbf{U}_p)+\Phi D\mathbf{U}_p + \Gamma\mathbf{U}_f \\
	&= \begin{bmatrix}\Phi (D+M\Gamma) & -\Phi M & \Gamma\end{bmatrix} \underbrace{\begin{bmatrix}\mathbf{U}_p^T & \mathbf{Y}_p^T & \mathbf{U}_f^T \end{bmatrix}^T}_{W}.
	\end{aligned}
	\end{equation*}
	For given $W$ and $\mathbf{Y}_f$ we can obtain the least--squares estimated prediction matrix $\Gamma$:
	\begin{equation}\label{eq:ls_markov}
	\begin{bmatrix}P_1 & P_2 & \Gamma\end{bmatrix} = \mathbf{Y}_fW^\dagger,
	\end{equation}
	where $[.]^\dagger$ represents the Moore-Penrose pseudo-inverse, $P_1 = \Phi (D+M\Gamma)$ and $P_2 = -\Phi M$. Hereby it should be noted that $L \geq (N)(2m+q)$ to ensure \eqref{eq:ls_markov} has a solution. 
	
\subsubsection{Estimating $\Phi$: measurable state}\label{sssec:MeasStateSeq} 
    If the system state is measurable, an estimate of $\Phi$ can be accurately obtained from data. The following ways to estimate it are derived from \cite{Goncalves&Bazanella&Lorenzini&Campestrini:2018}. The first one uses the estimated $\hat{\Gamma}$ and \eqref{eq:Hrel}, as follows:
	\begin{equation}
	\label{eq:Ometh1}
	\mathbf{Y}_p = \Phi X_p + \Gamma\mathbf{U}_p, ~~\text{which yields}~~	\Phi = (\mathbf{Y}_p-\hat{\Gamma}\mathbf{U}_p)X_p^\dagger.
	\end{equation}
	Another approach is using the orthogonal complement of the row space of the matrix $\mathbf{U}_p$:
	\begin{equation}
	\begin{aligned}
	\label{eq:Ometh2}
	\mathbf{U}_{po} & := I - \mathbf{U}_p^T(\mathbf{U}_p\mathbf{U}_p^T)^{-1}\mathbf{U}_p \\
	\mathbf{Y}_p\mathbf{U}_{po} &= \Phi X_p\mathbf{U}_{po}+\Gamma\mathbf{U}_p\mathbf{U}_{po} = \Phi X_p\mathbf{U}_{po}, \\
	\Phi &= (\mathbf{Y}_p\mathbf{U}_{po})(X_p\mathbf{U}_{po})^\dagger.
	\end{aligned}
	\end{equation}
	
\begin{remark}{\emph{(The LQR equivalence)}} When the state is measurable, one can also retrieve the stability  and (sub--)optimal performance properties of the LQR controller even for a small horizon $N$. Consider the following LQR and MPC cost functions, respectively:
\begin{align*}
J(y(k), U_k)&=\sum_{i=0}^{\infty}(y_{i|k}^{T}Qy_{i|k} + u_{i|k}^{T}Ru_{i|k}) \\
J(y(k), U_k) &=x_{N|k}^ TQ_px_{N|k} + \sum_{i=0}^{N-1}(y_{i|k}^{T}Qy_{i|k} + u_{i|k}^{T}Ru_{i|k})
\end{align*}
Notice that here $Q_p$ is a terminal penalty on the \textit{state vector}, not on the output. If $Q_p$ equals the solution of the associated DARE for the LQR problem, then we obtain the infinite horizon LQR control law as the unconstrained MPC control law\footnote{For sufficient large $N$ there exists a closed-form solution for $Q_p$ (see, e.g., \cite{Goncalves&Bazanella&Lorenzini&Campestrini:2018}), but then the MPC cost function also converges to the LQR case.}.  

In order to introduce the terminal state penalty in the developed data--driven predictive controller, one can create an unobservable output $y_o=Vx$, where $V$ is the Cholesky decomposition of $Q_p$, such that $y_o^Ty_o=x^TQ_px$. Thus, we have an augmented output vector $y_a=[y^T~~ y_o^T]^T$ which needs only to be re-substituted in \eqref{eq:hankel_data}. This also yields the following performance matrix changes:
$\tilde{Q}=\text{diag}(Q,\mathbf{0}_n),$
$\tilde{P}=\text{diag}(\mathbf{0}_q,\mathbf{1}_n).$
For sub-optimality it suffices to choose a $\tilde{Q}_p > Q_p$.
\end{remark}
	
\subsubsection{Estimating $\Phi x(k)$: only I/O data available}\label{sssec:OnlineUpdate}
    If the state is not measurable, the method of \cite{LQRpredict} can be used to estimate $\Phi x(k)$ on--line, using previous inputs and measured outputs, as follows:
	\begin{equation}
	\begin{aligned}
	\label{eq:Ometh4}
	\Phi x(k) &= \Phi A^{N}x(k-N) + \Phi D\mathbf{u}(k-N) \\
	&= -\Phi M\Phi x(k-N) + \Phi D\mathbf{u}(k-N)\\
	&= -\Phi M\mathbf{y}(k-N+1) + \Phi (D+M\Gamma)\mathbf{u}(k-N).
	\end{aligned}
	\end{equation}
	Since both $\Phi(D+M\Gamma)$ and $-\Phi M$ can be directly extracted from the matrix obtained in equation \eqref{eq:ls_markov}, then one can compute $\Phi x(k)$ in real--time as
	\begin{equation}
	\label{eq:Ometh41}
	\Phi x(k) = \begin{bmatrix}P_1 & P_2\end{bmatrix}\begin{bmatrix}\mathbf{u}(k-N)  \\ \mathbf{y}(k-N+1)\end{bmatrix}.
	\end{equation}
On--line estimation of $\Phi x(k)$ depends on the unknown $M$ and the estimate accuracy is not guaranteed to improve for any sequence length shorter than $N$. 

\begin{remark}{\emph{(Initial inputs and feasibility)}}
If $\Phi x(k)$ is computed using previous I/O data, it should be noted that due to this poor initial estimation, the predictive control quadratic program may not be feasible at start. This can be circumvented \emph{via} soft--constraints.
\end{remark}

\subsection{Integral action}\label{ssec:integral}
Embedding integral action within a model--based predictive controller is known to be beneficial both with respect to tracking time--varying references and removing off--sets. To this end, the rate--based MPC formulation was developed (see, e.g., \cite{Wang:2009}) by defining an augmented state $x_I(k) := \begin{bmatrix}\Delta x(k)^T & y(k)^T\end{bmatrix}^T$, where $\Delta x(k) := x(k) - x(k-1)$ and  an incremental input $\Delta u(k) := u(k) - u(k-1)$.
Then we can construct an augmented system model as follows:
\begin{equation}
\begin{aligned}
\label{eq:IntMod}
x_I(k+1) &= \underbrace{\begin{bmatrix}A & 0 \\ CA & I \end{bmatrix}}_{A_I}x_I(k) + \underbrace{\begin{bmatrix}B \\ CB\end{bmatrix}}_{B_I}\Delta u(k)\\
y(k) &= \underbrace{\begin{bmatrix}0 & I\end{bmatrix}}_{C_I}x_I(k).
\end{aligned}
\end{equation}
The cost function changes to:
\begin{equation*} \label{eq:IntCostFcn}
\begin{aligned}
J(x(k), \Delta U_k) & := (y_{N|k}-r_{N|k})^ TP(y_{N|k}-r_{N|k}) \\  
+ \sum_{i=0}^{N-1} & \left((y_{i|k}-r_{i|k})^{T}Q(y_{i|k}-r_{i|k}) + \Delta u_{i|k}^{T}R\Delta u_{i|k}\right) \\
& = \Delta U_k^T(\Psi+\Gamma_I^T\Omega\Gamma_I)\Delta U_k\\ 
&~~ + 2\Delta U_k^T\Gamma_I^T\Omega(\Phi_I x_I(k)-\mathcal{R}_k) \\
&~~  + (y(k) - r(k))^TQ(y(k) - r(k)) \\
&~~ +(\Phi_I x_I(k)-\mathcal{R}_k)^T\Omega(\Phi_I x_I(k)-\mathcal{R}_k).
\end{aligned}
\end{equation*}
where $\Delta U_k=\{\Delta u_{0|k}, \ldots,\Delta u_{N-1|k}\}$, $r_{0|k}=r(k)$ and $\mathcal{R}_k = \begin{bmatrix}r_{1|k}^T & r_{2|k}^T & \dddot{} & r_{N|k}^T\end{bmatrix}^T$ is a known future reference. The actual control input is obtained as $u(k) = u(k-1) + \Delta u^*(k)$.
By substituting the integral model \eqref{eq:IntMod} into the definitions in equation \eqref{eq:MatDef}, we obtain
\begin{equation}
\begin{aligned}
\label{eq:PI2}
\Phi_I & = \begin{bmatrix}CA & I \\ CA^2 + CA & I \\ \vdots \\ \sum_{i=1}^{N}CA^i & I\end{bmatrix}, \\ 
\Gamma_I & = \begin{bmatrix} CB & 0 & \dddot{} & 0 \\ CAB+CB & CB & \dddot{} & 0 \\ \vdots & \vdots & \ddots & \vdots \\  \sum_{i=0}^{N-1}CA^{i}B & \sum_{i=0}^{N-2}CA^{i}B & \dddot{} & CB \end{bmatrix}.
\end{aligned}
\end{equation}
We now formulate a method that allows us to derive these integral action prediction matrices directly from data by manipulating the input used in the experiment. 

\begin{theorem}\label{thm:int}
Let $[U_I]_{i|k} = \begin{bmatrix}u_{0|k} \\ u_{0|k}+u_{1|k} \\ \vdots \\ \sum_{j=0}^{i}u_{j|k}\end{bmatrix} + \begin{bmatrix}1 \\ 1 \\ \vdots \\ 1\end{bmatrix}\otimes\underbrace{\left(\sum_{j=0}^{k-1}u(j)\right)}_{u_{I}(k-1)},$
and $[\Phi]_i$, $[\Gamma]_i$ (respectively $[\Phi_I]_i$, $[\Gamma_I]_i$) represent the corresponding row-block w.r.t. output $y_{i|k}$. Then it holds that
\begin{equation}
\begin{aligned}
\label{eq:TheoremInt}
y_{i|k} &= [\Phi_I]_i [x_{I}]_{0|k} + [\Gamma_I]_i U_{i|k} \\
&=[\Phi]_i x_{0|k} + [\Gamma]_i [U_{I}]_{i|k}, ~~\forall i=1, \dddot{}, N.
\end{aligned}
\end{equation}
\end{theorem}

\begin{proof} 
Consider first the trivial case, $i=1$, and let $y_{0|k}=CAx(k-1)+CBu_I(k-1)$:
\begin{equation}
\begin{aligned}
\label{eq:DI_1}
y_{1|k} &= \begin{bmatrix}CA & I \end{bmatrix}[x_I]_{0|k} +  CB u_{0|k}\\
		 &= CAx_{0|k} - CAx(k-1) + y_{0|k} + CBu_{0|k}\\
     &= CAx_{0|k} - CAx(k-1) + CAx(k-1)\\
		 &~~~~ + CBu_I(k-1) + CBu_{0|k}\\
     &= CAx_{0|k}  + CB(u_I(k-1) + u_{0|k})\\
		 &=[\Phi]_1x_{0|k}+[\Gamma]_1[U_I]_{1|k}.
\end{aligned}
\end{equation}
Next, consider the case $i=2$, and the previous relation for $y_{0|k}$:
\begin{equation}
\begin{aligned}
\label{eq:DI1}
y_{2|k} & = \begin{bmatrix}CA^2 + CA\ & I \end{bmatrix}[x_I]_{0|k}\\
	& ~~~~ + \begin{bmatrix}CAB+CB & CB \end{bmatrix}\begin{bmatrix}u_{0|k} \\ u_{1|k}\end{bmatrix} \\
    &= (CA^2 + CA)(Ax(k-1)+Bu_{I}(k-1))\\ 
			&~~~~- (CA^2 + CA)x(k-1) + CAx(k-1) \\ & ~~~~ + CBu_{I}(k-1)  + (CAB+CB)u_{0|k} + CBu_{1|k}\\
    &= CA^2(Ax(k-1)+Bu_{I}(k-1)) \\ 
		& ~~~~ +(CAB+CB)(u_{I}(k-1)+u_{0|k}) + CBu_{1|k}\\
    &= CA^2{x}_{0|k} + \begin{bmatrix}CAB & CB \end{bmatrix}\begin{bmatrix}u_{0|k}+u_{I}(k-1) \\  u_{1|k}+u_{0|k}+ u_{I}(k-1)\end{bmatrix}\\
		& =[\Phi]_2x_{0|k}+[\Gamma]_2[U_I]_{2|k}.
\end{aligned}
\end{equation}
Thus, we obtained the prediction of output as a function of the original prediction matrices, but for the integrated input. The proof for any other $i\geq 3$ follows using the same derivations and is omitted due to space limitation.
\end{proof}

The result of Theorem~\ref{thm:int} suggests that it is possible to estimate the prediction matrices corresponding to the rate--based predictive control algorithm  by manipulating the data fed to the system in the experiment. More specifically, the following procedure can be applied. 
First, create an input set $U$ and form a second input set $U_I$ from $U$ such that for any $k\geq 0$, $U_I(k) =\sum_{i=0}^{k}U(i)$. Next, perform the experiment on the system using $U_I$ as input and collect the output, which we will denote $Y_I$. Finally, compute $\Gamma_I$ as in \eqref{eq:ls_markov} using $Y_I$ for $\mathbf{Y}_p$ and $\mathbf{Y}_f$, but using $U$ for the input matrices $\mathbf{U}_p$ and $\mathbf{U}_f$, instead of the applied $U_I$.
Using $U_I$ in the experiment, however, could lead to a high magnitude input. An alternative procedure can be used to circumvent this problem, as follows: apply $U$ in the experiment, collect $Y$ and use $U_{\Delta}(k) = U(k) - U(k-1)$ to compose $\mathbf{U}_p$ and $\mathbf{U}_f$. 
The proof that the developed estimation procedure can still be applied follows the same derivations as above, but we do not include it here due to space limitations. 

\begin{remark} The implementation of the integral action DPC 
still requires estimation of $\Phi_I x_I(k)$. If the states are measurable (see Section~\ref{sssec:MeasStateSeq}) the unconstrained integral DPC law becomes:
\begin{equation}
\begin{aligned}
\label{eq:IntImp1}
\Delta U_k^*(x(k)) &= -G^{-1}F\left(\Phi_I \begin{bmatrix}x(k)-x(k-1) \\ y(k)\end{bmatrix} - \mathcal{R}_k\right).
\end{aligned}
\end{equation}
To obtain an estimate of $\Phi_I$, equations \eqref{eq:Ometh1} or \eqref{eq:Ometh2} can be used in combination with the state vector corresponding to the integral action augmented state, i.e., $x_I(k) = [\Delta x(k)^T y(k)^T]^T$. If the state is not measurable, the unconstrained integral DPC law changes to 
\begin{align*}
\Delta U_k^*(x(k)) &= -G^{-1}F\left(\begin{bmatrix}P_1 & P_2\end{bmatrix}\begin{bmatrix}\Delta\mathbf{u}(k-N) \\ \mathbf{y}(k-N+1)\end{bmatrix}- \mathcal{R}_k\right)\\
&=-G^{-1}F\left(\Phi_I x_I(k)- \mathcal{R}_k\right),
\end{align*}
which is similar to the method used in Section~\ref{sssec:OnlineUpdate}. If constraints are added to the optimization problem, the same approaches to estimate $\Phi_I$ or $\Phi_I x_I(k)$, respectively, apply, as the constraints matrices \eqref{eq:constr2} (corresponding to the integral action formulation) only depend on the estimated prediction matrices.
\end{remark}

\section{Illustrative examples}

\subsection{Position control for a linear motor} \label{sec:4}
In this section we provide an illustrative example of the proposed methodology for the position control of a linear motor, see, e.g., \cite{Tuan:2018}. The simplified motion dynamics of the mechanical part of the actuator can be represented by the continuous--time model
$$\dot{x}(t) = \begin{bmatrix} 0 & 1 \\ 0 & -50 \end{bmatrix}x(t) + \begin{bmatrix} 0 \\ 0.5 \end{bmatrix} u(t) + \begin{bmatrix} 0 \\ 0.5 \end{bmatrix} d(t),$$
$$y(t) = \begin{bmatrix} 1 & 0 \end{bmatrix} x(t),$$
where the first state (the output) is the position and the second state is the velocity. The system is discretized with a sampling time of $T_s = 0.02$~s. In order collect data from the system we set an experiment where the input is a PRBS signal with an amplitude of $\pm 20$~N  at $50$~Hz and length 6022 samples. The output is corrupted by a white noise sequence such that the Signal-to-Noise Ratio (SNR) is $26$~dB.

The control performance parameters are given by $N = 10$, $Q = 6\text{e}5$, and $R = 0.005$. Also the input force is constrained between $-500~\text{N}\leq u(k) \leq 500~\text{N}$, and  displacement is limited between $-0.165~\text{m} \leq y(k) \leq 0.165~\text{m}$. We then perform the estimation of the prediction matrices in \eqref{eq:ls_markov} and \eqref{eq:Ometh2} with $L=3000$, and similarly for the integral DPC case. 

\begin{figure}[h]
\centering
\includegraphics[width=0.5\textwidth, trim={0.1cm 2.3cm 2cm 0.5cm},clip]{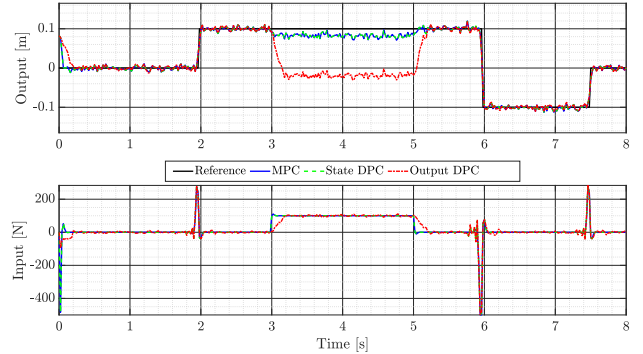} 
\caption{Linear motor comparison of 3 predictive controllers: standard MPC; state DPC (with measured state); output DPC (with measured output).}
\label{fig:MPC1}
\end{figure}

Fig.~\ref{fig:MPC1} portrays the closed--loop results for a sequence of steps using 3 predictive controllers: standard model--based PC, the developed DPC with measured state and  measured output, respectively. Also, between $t=3$~s and $t=5$~s we apply a constant disturbance force of amplitude $d(k) = -100$~N and we also consider the output corrupted by white noise with SNR = $25$~dB. It can be observed that the DPC with measured state converges faster to the reference compared to the DPC with measured output, and it matches the standard MPC result. Due to the poor initial estimation of the output DPC, the initial response is a bit slower. When the disturbance becomes active all 3 predictive controllers result in an off--set, which is considerably larger for the output DPC.

\begin{figure}[h]
\centering
\includegraphics[width=0.5\textwidth, trim={0.1cm 2.3cm 2cm 0.5cm},clip]{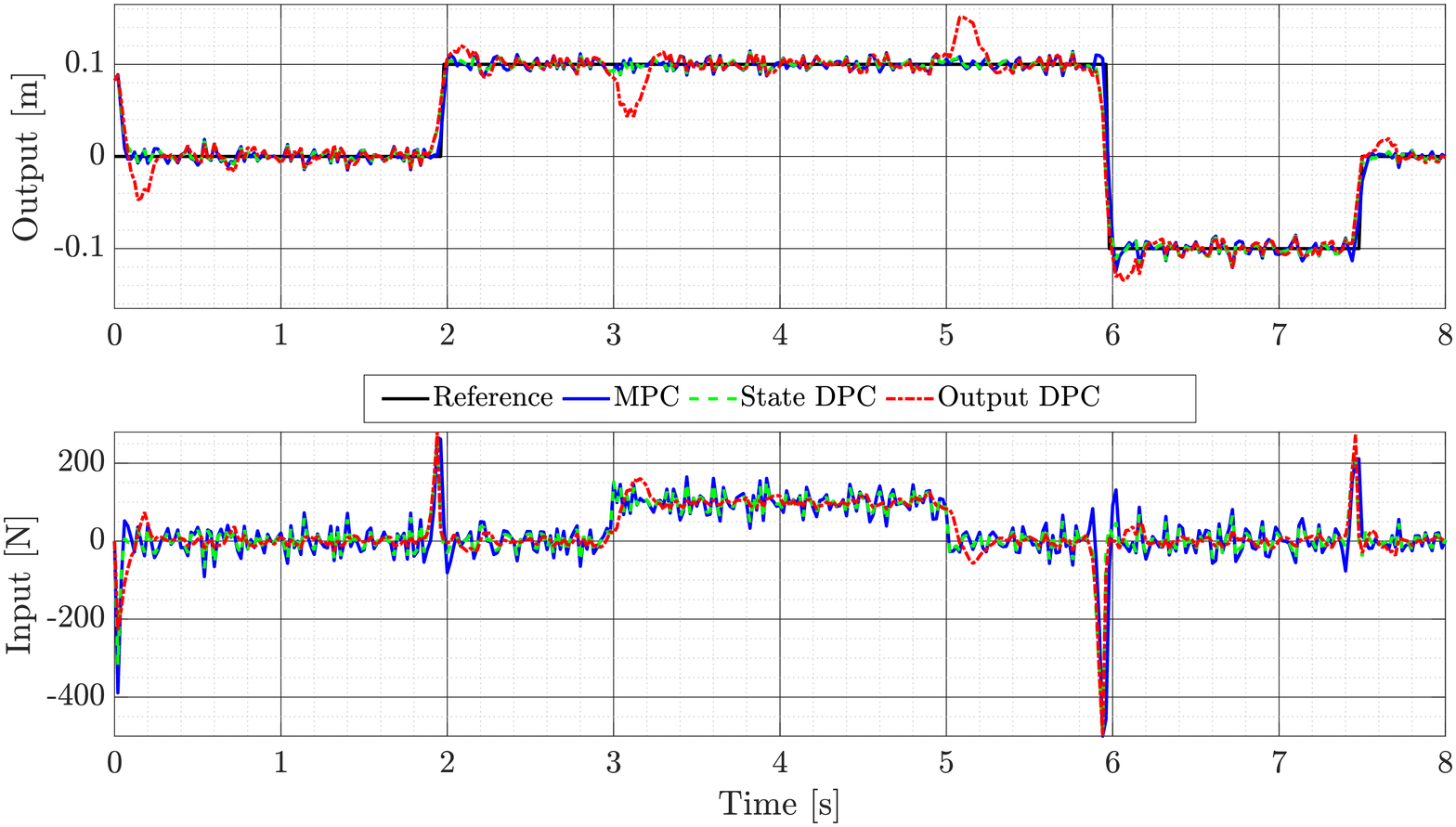} 
\caption{Linear motor comparison of 3 integral predictive controllers: standard i--MPC; state i--DPC (with measured state); output i--DPC (with measured output).}
\label{fig:MPC2}
\end{figure}

Fig. \ref{fig:MPC2} shows the closed--loop simulation results for the corresponding 3 integral predictive controllers. All controllers successfully remove the off-set caused by the disturbance, while the output integral DPC exhibits longer transients due to the on--line estimation part. This really demonstrates the need of incorporating integral action in data--driven predictive controllers and also, the effectiveness of the developed DPC algorithm. 

\subsection{UPS system with periodic disturbance} \label{ssec:SimCompInt2}
\begin{figure}[h!]
	\centering
	\includegraphics[width=0.49\textwidth, trim={1cm 2.2cm 3cm 0.3cm},clip]{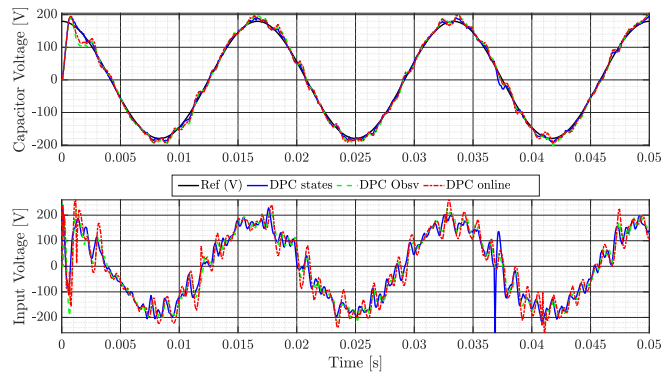}
	\caption{Comparison of Data--Driven PC methods on an UPS system}
	\label{fig:UPS2}
\end{figure}

To provide another practical example, the data--driven control algorithm with integral action is also applied to an uninterruptible power supply (UPS). This plant has been studied before in \cite{Goncalves&Bazanella&Lorenzini&Campestrini:2018}.
\begin{figure}[h]
	\centering
	\includegraphics[width=0.49\textwidth, trim={1.5cm 1.2cm 0.7cm 0.8cm},clip]{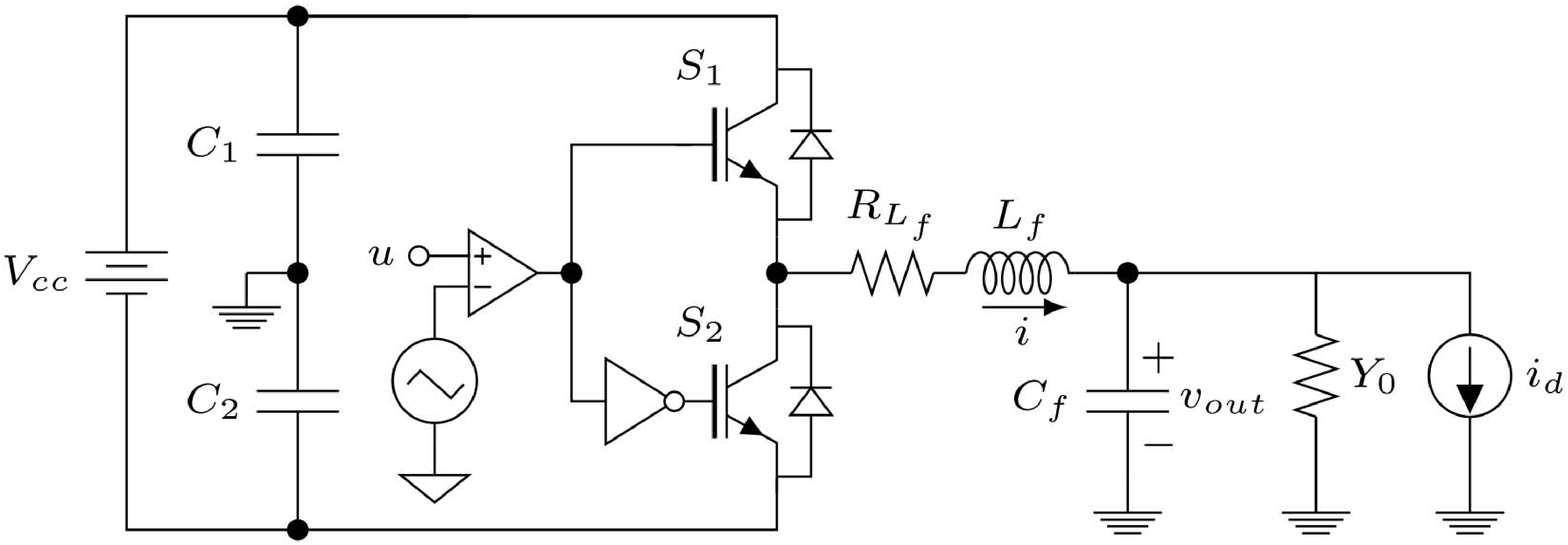}
	\caption{Simplified schematic of an UPS system}
	\label{fig:UPS}	
\end{figure}
Fig.~\ref{fig:UPS} shows a simplified electrical diagram of the output stage of a single-phase UPS system.
The load effect on the system output is modeled by a parallel connection of an uncertain admittance $Y_0(t)$ and an unknown periodic disturbance given by the current source $i_d(t)$. Again for simplicity, the uncertain admittance $Y_0(t)$ is presumed to be constant at $0.1519\Omega$. The input is a PWM control voltage and the output is the voltage over the capacitor. The states are the inductor current and the capacitor voltage, $x(t) = \begin{bmatrix} i(t) & v(t)\end{bmatrix}^T$. The system can be represented as the following state-space model:
$$\dot{x}(t) = \begin{bmatrix} -\frac{R_{Lf}}{L_f} & -\frac{1}{L_f} \\ \frac{1}{C_f} & -\frac{Y_0}{C_f} \end{bmatrix}x(t) + \begin{bmatrix} \frac{K_{PWM}}{L_f} \\ 0 \end{bmatrix} u(t) + \begin{bmatrix} 0 \\ \frac{-1}{C_f} \end{bmatrix} i_d(t) $$

\begin{equation}
\begin{aligned}
\dot{x}(t) &= \begin{bmatrix} -15 & -1000 \\ \frac{1}{3e-4} & -506.46 \end{bmatrix}x(t) + \begin{bmatrix} 1000 \\ 0 \end{bmatrix} u(t) + \begin{bmatrix} 0 \\ \frac{-1}{3e-4} \end{bmatrix}i_d(t) \\
y(t) &= \begin{bmatrix} 1 & 0 \end{bmatrix} x(t)
\end{aligned}
\end{equation}
The system is discretized with a sampling time of $T_s = 1/15000~s$ using a zero-order hold. The input voltage is constrained between $-260\leq u(t) \leq 260$ and there are no output constraints. The closed--loop is set to follow a sinusoidal reference $r(t) = 127\sqrt{2}\text{cos}(120\pi t)$.
Additionally, the following control parameters are used:
$$Q = 200, ~~ R = 50, ~~ N=15, ~~ L=3000$$
The system is measured off--line using a PRBS with an amplitude of $\pm104V$ with a length $7500$ samples at $15000$Hz. The experiment is corrupted with output noise such that the SNR is $28$dB, the disturbance current was not present in the experiment. During the simulation, the system is also subjected to an output noise with a SNR of $36$dB. The periodic disturbance is defined as multi--sine with 12 components with random frequencies between $2000$ and $12000$ Hz and has an amplitude of $\pm20A$.
See Fig.~\ref{fig:UPS2} for the result of the simulation. The various controllers all achieve reference tracking even for a periodic reference. The total harmonic distortion (THD) caused by the periodic disturbance on the system is between 3.8\% and 4\%.

\section{Conclusion}\label{sec:5}
In this work we proposed a new data--driven predictive control approach with estimated prediction matrices and integral action. Our approach lies between the direct data--enabled predictive control and the identified model--based predictive control. In this way, we are able to avoid identification of a complete state--space model and the respective system order, while still enjoying standard properties of least--squares estimation. Also, comparatively to the data--enabled approach, our method deals well with noisy data and requires less in general computations for the on--line implementation.  We also presented a formulation that allows direct incorporation of integral action in the DPC algorithm by only manipulating data that is fed to algorithm, without altering the single experiment required for estimation. Simulation results for position control of a linear motor illustrated the effectiveness of our methodology in terms of tracking performance, robustness to noisy measurements and disturbance rejection.


\bibliographystyle{IEEEtran}
\bibliography{baseref.bib}

\end{document}